# Les indicateurs avancés de l'inflation en RD. Congo

Henry N. MUGANZA[1]

*Abstract*

*This study aims to identify the leading of inflation indicators of monetary policy in DRC. The results reveal that the most relevant inflation indicators usually come from the monetary origin than the real sector. Variance decomposition analyzes place in the foreground the rate of exchange, the money supply and the public consumption like the most relevant indicators. In order to achieve its goal of price stability and to support a strong economic growth, the intermediate objective of the Central Bank baseded on the controle of the money supply seems to be less relevant. Relates to a high level of the dollarization, the central bank will be able to adopt either the strategy of nominal anchoring of the rate of exchange, this calls the return of the fixed exchange regime or to adopt a strategy of inflation targeting is to restore the credibility of the monetary policy*

## 1. Introduction

Le débat sur les indicateurs avancés d'inflation est au cœur de l'analyse monétaire depuis les années 90 avec l'adoption du régime de ciblage d'inflation comme cadre de référence dans la conduite de la politique monétaire (L. LEIDERMAN et L. SVENSSON, 1995 ; B. BERNANKE et F. MISHKIN, 1997).

Par indicateurs avancés d'inflation, il faut entendre un ensemble de variables économique et financière qui ont pour fonction de transmettre à la banque centrale les informations sur les variations actuelles et futures du niveau général des prix (C. FREEDMAN, 1996 ; J. P. ALLEGRET et J. F. GOUX, 2003 ; B. LANDAIS, 2008).

Dans ce papier, le débat porte essentiellement sur la définition à court terme du mécanisme d'ancrage nominal permettant d'orienter les anticipations des agents privés sur une trajectoire compatible avec l'objectif fixé[2]. A ce sujet, deux types d'approches peuvent être distinguées : 1°) le ciblage d'agrégats monétaires ; 2°) le ciblage direct de l'inflation (G. LEVIEUGE, 2003 ; T. BRAND, 2008).

---

[1] Assistant de recherche et d'enseignements à la Faculté des Sciences Economiques et de Gestion ; Université Evangélique en Afrique. Mail : henryngongo06@gmail.com

[2] Un consensus s'est établi sur l'engagement de la banque centrale à stabiliser les prix, c'est-à-dire maintenir un niveau faible et stable d'inflation, l'annoncer au public sous forme d'une cible ponctuelle ou à travers une fourchette cible afin de promouvoir le bien-être économique. En agissant ainsi, la banque centrale réduit le risque d'incohérence temporelle et accroit la crédibilité de la politique monétaire (J. P. POLLIN, 2008).

En ce qui concerne la première approche, le consensus sur la politique monétaire forgé par M. FRIEDMAN (1968) visait à contrôler les agrégats monétaires pour atteindre la stabilité des prix. Cette formulation monétariste fondée sur une norme de croissance des agrégats monétaires repose sur l'hypothèse d'exogénéité totale de la monnaie, d'une part, et sur un certain nombre de critères, à savoir : le lien de causalité stable entre l'agrégat de monnaie choisi comme objectif et les prix, c'est-à-dire l'existence d'une fonction de demande de monnaie stable, d'autre part. L'annonce du taux de croissance de la masse monétaire répond principalement à un souci de crédibilité afin de peser sur les anticipations inflationnistes. Ce canal fait donc de l'inflation un phénomène monétaire (A. BENASSY et al, 2005 ; P. R. AGENOR, 2008 ; S. ENGONE MVE, 2010).

De ce point de vue, le ciblage monétaire a été adopté par de nombreux pays au cours des années 70 jusqu'au début des années 80. Cependant, le contrôle de la base monétaire, c'est-à-dire de la quantité de monnaie émise par la banque centrale n'est pas parvenu à assurer la maîtrise de l'inflation sur le long terme. Le socle du fondement monétariste a été rendu caduc en partie par l'instabilité des fonctions de demande de monnaie engendrée par la libéralisation financière des années 80. Ainsi, la remise en cause du caractère exogène de la monnaie et du lien de causalité stable entre les agrégats monétaires et l'inflation a fait perdre à cet agrégat son statut d'objectif intermédiaire et même en tant qu'indicateur dans certains pays. C'est ce qui a conduit au recours au taux d'intérêt comme variable intermédiaire conformément à l'enseignement de Poole (1970)[3] (BLUNDELL- WIGNALL et al, 1990 ; J. P. POLLIN, 2004 ; S. ENGONE MVE, 2010).

En effet, comme le taux d'intérêt ne peut à lui seul rendre compte des effets de la politique monétaire dans une économie ouverte, certaines banques centrales, en particulier celles du Canada et de la Nouvelle Zélande, ont préféré tenir compte de l'indice des conditions monétaires dans le cadre d'un pilotage plus direct de la cible final. Cet indice associe à la fois le taux de change et le taux d'intérêt en fonction de leur impact relatif sur la croissance et l'inflation. Cet indice synthétique peut alors être utilisé, soit à des fins d'analyse conjoncturelle, soit en tant que simple indicateur de politique monétaire. Il doit à cet effet justifier de bonnes capacités prédictives si bien que les banques centrales qui s'y réfèrent pour justifier une recommandation de politique monétaire ou fonder une analyse conjoncturelle devraient le considérer avec une certaine prudence (S. DIARISSO et O. SAMBA MAMADOU, 2000 ; L. AUBERT, 2003 ; C. BLOT et G. LEVIEUGE, 2008).

De façon générale, la libéralisation financière a battu en brèche les règles monétaires passives à la Friedman. Ainsi, le contrôle des agrégats monétaires tout comme du taux de change[4], en tant qu'objectifs intermédiaires, ne permet pas de prendre en compte les évolutions de l'économie et sont par conséquent qualifiées de sous-optimale pour atteindre la stabilité des prix. D'où, l'apparition depuis les années 90 d'une conception beaucoup plus pragmatique de la formulation fondée sur les prévisions du taux d'inflation comme cible idéale de la politique monétaire (BLUNDELL- WIGNALL et al, 1990 ; J. CREEL et H. STERDYNIAK, 1999 ; J. P. POLLIN, 2005 ; E. SVENSSON, 1997).

---

[3] Cet auteur fait valoir qu'en règle générale, l'économie subie des chocs transitoires aussi bien dans le secteur réel que dans le secteur monétaire. Il affirme alors que lorsque la perturbation provient du marché de la monnaie, la politique monétaire doit contrôler le taux d'intérêt et lorsqu'elle a pour origine le marché des biens, on doit se fixer un objectif de croissance monétaire.

[4] Dans le contexte du régime de change fixe, le taux de change a été utilisé par un certain nombre de pays émergents comme objectif intermédiaire de la politique monétaire. Mais, cette cible a perdu du terrain en raison de conflits générateurs de crises entre les objectifs des politiques internes et externes des banques centrales. Elle constitue une perte d'indépendance pour la politique monétaire du pays qui a décidé d'ancrer sa monnaie à celle d'un autre pays, encourage la spéculation et fragilise le système financier.

La deuxième approche quant à elle repose sur l'idée d'une monnaie endogène et est centrée sur la question du ciblage direct d'inflation comme stratégie efficace pour atteindre la stabilité des prix. Cette approche a été suggérée par un certain nombre d'auteurs, entre autres L. LEIDERMAN et L. SVENSSON (1995), L. SVENSSON (1997, 1998, 1999), BERNANKE et MISHKIN (1997), à la suite de l'abandon progressif du contrôle des agrégats monétaires au cours des années 80 (S. ENGONE MVE, 2010).

La mise en œuvre de cette stratégie soulève toutefois des controverses entre les partisans de la règle d'instrument et les tenants de la règle d'objectif. Cette opposition est illustrée par le débat entre B. McCALLUM et E. NELSON d'une part et L. SVENSSON d'autre part (B. LANDAIS, 2008).

Si l'on en croit les partisans de la règle d'instrument, B. McCALLUM (1987) et J. B. TAYLOR (1993), les règles ont des avantages majeurs sur les politiques discrétionnaires dans la mesure où les décisions prises dans un environnement non discrétionnaires ont des effets favorables sur les anticipations des agents.

Ainsi, convaincu des bienfaits d'une règle monétaire, la règle de McCALLUM (1987, 1995) stipule que le rythme d'ajustement de la base monétaire est fonction de l'écart entre le PIB nominal et sa cible et/ou en cas de variation permanente de la vitesse de circulation de la monnaie. Cette règle repose sur la cible de PIB nominal et sur l'utilisation de la base monétaire comme instrument de la politique monétaire (J. J. DURAND et N. PAYELLE, 1998).

De même que B. McCALLUM, J. TAYLOR (1993) soutient l'importance des règles dans la conduite de la politique monétaire. Cependant, il estime que c'est le taux d'intérêt qui doit être retenu comme instrument de la politique monétaire et non pas la base monétaire. Tout en mettant en évidence le rapport incertain qui peut exister entre la base monétaire et l'inflation, la règle de TAYLOR fait valoir qu'il est commode pour la banque centrale d'ajuster son taux directeur en fonction des écarts de l'inflation par rapport à sa cible, des excès de l'output par rapport à son niveau potentiel[5].

A en croire, au contraire, les tenants de la règle d'objectif, la banque centrale ne cible pas les variables intermédiaires, mais elle agit directement sur l'objectif d'inflation. Faute d'un consensus sur un canal de transmission prédominant à travers lequel la politique monétaire opère au sein de l'économie, c'est la prévision de l'inflation qui tient lien d'objectif intermédiaire. La banque centrale doit donc surveiller l'évolution d'un certain nombre d'indicateurs reconnus comme étant aptes à prévoir l'inflation pour parvenir à contrecarrer les tensions inflationnistes avant qu'elles ne se concrétisent. L'intervention de la banque centrale sur le marché monétaire vise dans ce cas à ramener le taux d'inflation anticipée proche de l'inflation ciblée (L. SVENSSON, 1997 ; P. R. AGENOR, 2008 ; L. GREGORY, 2003).

---

[5] Simples soient-elles, les règles d'instrument soulèvent un certain nombre de critique entre autres le fait de ne tenir compte que d'un trop petit nombre d'indicateurs dans l'orientation des actions monétaires. Aussi, outre la critique de Lucas, l'économie ne réagit qu'aux variations de l'output ou de l'inflation en suivant des règles de version *Backward-Looking*. En présence d'autres types de choc, la banque centrale qui s'appui sur un tel modèle ne peut pas aboutir à son objectif de stabilité des prix. Pour le tenant de la règle d'objectif, ces règles, comme l'approche par les objectifs intermédiaires, sont qualifiées de sous optimal pour atteindre l'objectif fixé (A. SIRRI, 2007 ; Z. FTITI et J. F. GOUX, 2011).

Cette approche est formalisée selon l'expression de B. BERNANKE et F. MISHKIN (1997) comme un cadre de « discrétion contrainte »[6] qui laisse aux autorités monétaires la latitude d'utiliser toutes les informations ou indicateurs qu'elles jugent pertinentes pour connaître l'orientation des anticipations des agents et pouvoir ainsi décider de l'orientation future de la politique monétaire. Ces informations comportent plusieurs indicateurs, financiers et réels, susceptibles de rendre compte des origines multiples de l'inflation et d'éclairer les autorités monétaires sur l'évolution future de l'inflation (F. MISHKIN, 2004 ; S. CECCHETTI et al, 2000 ; N. PAYELLE et al, 2001; T. BRAND, 2008 ; B. DIANE, 2011).

Ainsi, l'objet de ce travail est d'apprécier la pertinence de l'usage des agrégats monétaires par la Banque Centrale du Congo[7] comme cible intermédiaire et Notre étude portera sur la R.D.Congo compte tenu de son histoire inflationniste spectaculaire et de son niveau élevé de dollarisation. Elle est particulièrement importante aujourd'hui pour au moins deux raisons :

1°) l'engagement de la Banque Centrale du Congo à maintenir un taux d'inflation faible et stable. Ce qui suscite de surcroît l'intérêt de déterminer l'indicateur d'inflation fondamentale qui est un dispositif nécessaire permettant à une banque centrale d'améliorer l'efficacité de son action ;

2°) la volonté manifestement exprimée depuis 2001, après quasiment plus de cinquante ans d'opacité, de renforcer la transparence pour conquérir sa crédibilité à travers la mise en œuvre d'un dispositif de communication avec le public.

De ce qui précède, notre analyse voudrait alors répondre à la question suivante : quels sont les indicateurs d'inflation pertinents pour la conduite de la politique monétaire en R.D.Congo ?

## 2. Methodologie

Il convient de souligner que nous allons utiliser un modèle économétrique dont les données seront analysées à l'aide du logiciel *Eviews 7*. Ainsi, dans ce point, nous tenons à présenter le modèle qui offre un cadre à partir duquel la question d'identification des indicateurs avancés d'inflation est analysée. Ainsi, l'objet de notre modèle est de déterminer les variables « indicateurs » qui peuvent prévenir l'évolution des prix et des menaces pour la stabilité des prix et dont la surveillance permet aux autorités monétaires de prendre des mesures appropriées sur l'orientation de la politique monétaire.

Pour identifier les indicateurs d'inflation, la littérature retient plusieurs types des modèles notamment les modèles VAR, les modèles factoriels dynamiques, les modèles structurels (Nouvelle courbe de Phillips Hybride, NKPC), les modèles P-star, les modèles DSGE, les modèles GPM, etc. (M. B. DIOP, 2011).

---

[6] C'est une disposition intermédiaire entre la discrétion absolue, porteuse d'incohérence temporelle et génératrice de biais d'inflation et le suivi aveugle de règles intangibles. A l'intérieur des contraintes imposées par les objectifs d'inflation de moyen et long terme, les banquiers centraux se sont accordés de grandes latitudes pour répondre aux conditions de l'emploi, des changes et autres développements de court terme (B. LANDAIS, 2008).

[7] Des lors que l'objectif intermédiaire, masse monétaire, crédit à l'économie ou taux de change, suivi par la banque centrale contient toute l'information utile à la prévision de l'inflation, utiliser un objectif intermédiaire est équivalent à cibler la prévision de l'inflation (B. S. BERNANKE et F. S. MISHKIN, 1997).

Notre choix va porter sur la modélisation VAR. Ce modèle a l'avantage non seulement d'intègrer un certain nombre d'indicateurs, mais encore il permet de bien identifier les capacités prédictives d'un indicateur à travers l'analyse de la décomposition de la variance. Il pose toutefois le problème de degré de liberté si bien que nous n'avons retenu qu'un nombre réduit de variables.

### 2.1. Modèle théorique

La présente étude s'appuie sur le modèle développé par J. H. STOCK et M. W. WATSON (1999) pour évaluer la pertinence des indicateurs d'inflation aux Etats-Unis entre 1959 et 1997. La forme simplifiée de ce modèle telle que synthétisé par S. KOZICKI (2001) se présente de la manière suivante :

$$P_t = f(Ti, M2, TCE, CHOM, TX) \qquad (1)$$

Où :

$P_t$ : Le taux d'inflation ; $Ti$ : le taux d'intérêt ; $M2$ : le taux de croissance de la masse monétaire ; $TCE$ : le taux de change ; $CHOM$ : le taux de chômage ; $TX$ : le taux de croissance du produit intérieur brut.

Dans le cadre de notre analyse, nous allons donc adapter ce modèle aux spécificités de la R.D.Congo. En effet, l'inflation serait en même temps influencée directement par la politique budgétaire, la politique monétaire, et indirectement, à moyen et long terme par le canal de la production nationale. Ainsi, nous allons utiliser l'inflation comme variable expliquée, d'une part et le déficit budgétaire, la masse monétaire, le taux de change et le niveau de production comme variables explicatives d'autre part. Ce qui nous conduit à présenter notre modèle à des fins d'estimation.

### 2.2. Modèle à des fins d'estimation

Notre modèle à des fins d'estimation inspiré de J. H. STOCK et M. W. WATSON (1999) tel que synthétisé par S. KOZICKI (2001) et adapté aux spécificités de la R.D.Congo peut prendre la forme linéaire suivante :

$$INFL_t = \alpha + \beta_1 DF_t + \beta_2 M2_t + \beta_3 TCE_t + \beta_4 PIB_t + \varepsilon_t \qquad (2)$$

où :

$INFL_t$ : Le taux d'inflation. Cet indicateur est utilisé par toutes les Banques Centrales pour définir la stabilité des prix en dépit des erreurs systématiques qu'il contient ; $DF_t$ : Le taux de croissance de la consommation publique utilisé comme proxy du déficit budgétaire afin de mieux mesurer l'impact de la politique budgétaire sur l'inflation ; $M2_t$ : La croissance de la masse monétaire. Cette variable est traditionnellement considérée comme un indicateur avancé de l'inflation conformément aux enseignements de la théorie quantitative de la monnaie ; $TCE_t$ : L'indice du taux de change. C'est un indicateur par lequel les chocs externes influent sur le niveau de l'inflation. Le suivi de cet indicateur dans une petite économie ouverte à régime de change flexible semble intéressant pour l'orientation de la politique monétaire ; $PIB_t$ : Le taux croissance de l'économie. Il agit positivement sur l'inflation d'une part par son action sur la demande de biens non échangeables et d'autre part (avec un délai sans doute plus long) par son action sur le chômage et la croissance des salaires nominaux.

En termes de signes attendus, il est logique de postuler que : $\beta_1$ (-/ +), $\beta_2$ (+), $\beta_3$ (+/-), $\beta_4$ (+): respectivement les coefficients par rapport aux variables dont elles sont associées et (.) les signes théoriques; $\varepsilon_t$ est le terme d'erreur.

Il est important de noter que le choix de ces indicateurs est beaucoup plus justifié par leurs caractères endogènes, c'est-à-dire les variables sur lesquelles les autorités nationales peuvent agir et dont la plupart sont considérées comme des cibles intermédiaires potentielles de la politique monétaire.

## 3. Resultats

La base de données que nous utilisons dans notre analyse économetrique est constituée à partir du CD-Rom de la Banque Mondiale, « Indicateurs sur le Developpement dans le Monde » (WDI), dans sa version 2012 et subsidiairement complétées par les données de la Banque Centrale du Congo. La période considerée s'etale entre 1976 – 2011.

Pour ne pas tomber dans le risque d'estimer des relations « fallacieuses » et d'interpréter les résultats de manière erronée, il s'avère nécessaire de passer par une analyse préliminaire des données c'est-à-dire proceder au test de stationarité[8] et de cointégration.

### 3.1. Test de cointégration de JOHANSEN

Pour tester la cointégration de nos séries, nous allons utiliser l'approche de Johansen (1988). En effet, cette approche permet d'identifier la relation d'équilibre de long terme entre deux ou plusieurs variables en recherchant l'existence d'un vecteur de cointégration, c'est-à-dire s'assurer de la convergence des sentiers de croissance des variables sur le long terme. Ci-dessous les résultats du test de cointégration.

*Tableau n° 1. Résumé du test de cointégration de Johansen*

Sample (adjusted): 1978 2011
Trend assumption: Linear deterministic trend (restricted)
Series: INF DF M2 TCE GDP
Lags interval (in first differences): 1 to 1
Unrestricted Cointegration Rank Test (Maximum Eigenvalue)

| Hypothesized No. of CE(s) | Eigenvalue | Max-Eigen Statistic | 0.05 Critical Value | Prob.** |
|---|---|---|---|---|
| None * | 0.804568 | 55.50648 | 38.33101 | 0.0002 |
| At most 1 | 0.563135 | 28.15646 | 32.11832 | 0.1413 |
| At most 2 * | 0.537256 | 26.19979 | 25.82321 | 0.0446 |
| At most 3 | 0.262113 | 10.33478 | 19.38704 | 0.5831 |
| At most 4 | 0.142316 | 5.219651 | 12.51798 | 0.5652 |

Max-eigenvalue test indicates 1 cointegrating eqn(s) at the 0.05 level
* denotes rejection of the hypothesis at the 0.05 level
**MacKinnon-Haug-Michelis (1999) p-values

*Sources : test effectué à l'aide du logiciel Eview 7*

---

[8] Les resultats du test de statinnarité de Dickey-Fuller Augmenté attestent que les variables sont toutes intégrer de même ordre I (1) c'est-à-dire en différence première. D'où l'existence d'un risque de cointégration. Ce qui nous a conduit à passer au test de cointégration.

Le test de cointégration effectué sur l'équation (3) indique la présence d'une seule relation d'équilibre à long terme entre la variable expliquée et les différentes variables explicatives parce qu'on obtient une seule valeur du test max-eigen supérieur aux valeurs critiques au seuil de 5%.

Si le test de cointégration permet de détecter la présence d'une relation de long terme entre les variables, il est aussi important de connaitre l'évolution à court et moyen terme de cette relation. L'outil nécessaire pour parvenir à une telle fin est le modèle à correction d'erreur dont l'objectif est d'éliminer l'effet de vecteur de cointégration, d'une part, et de rechercher la liaison réelle entre les variables, d'autre part.

Ainsi, le modèle d'identification des indicateurs avancés de l'inflation fait intervenir un mecanisme d'ajustement dynamique vers une cible de long terme. Les relations entre les variables de l'équation (3) peuvent être représentées à l'aide d'un modèle vectoriel à correction d'erreur sous la forme suivante :

$$\Delta X_t = \sum_{i=1}^{p-i} \beta_t \Delta X_{t-i} + \pi X_{t-i} + \varepsilon_t \qquad (3)$$

Avec :

$$X_t = (INFL_t, DEF_t, M2_t, TCE_t, GDP_t)^T$$

Soit le vecteur des variables où $T$ désigne la transposé de $X_t$ ;

$\pi X_{t-i}$ : désigne la dynamique de long terme. La matrice $\pi$ permet de décrire les effets de long terme. A partir de la procédure de JOHANSEN la matrice $\pi$ peut être réécrite sous la forme $\pi = \alpha \beta^T$ où la matrice $\alpha$ est la force de rappel vers l'équilibre, il doit être significatif et nécessairement compris entre -1 et 0. Elle mesure la vitesse d'ajustement aux équilibres de long terme et $\beta^T$ constitue le vecteur de cointégration. Il s'agit donc de la matrice dont les éléments sont les coefficients des relations de long terme des variables.

$\varepsilon_t$ : Vecteur des erreurs $\sim N(0,\Sigma)$ c'est-à-dire normalement distribuée;

$\Delta$ : Opérateur de différence première ; et

$\beta_i$ et $\pi$ désignent respectivement les matrices des coefficients de long terme et court terme.

Apres cette analyse préliminaire qui débouche sur la mise en évidence ou la présentation du modèle qui va nous servir de référence à des fins d'estimation, nous pouvons dès lors procéder à l'estimation du modèle.

### 3.2. Résultats de l'estimation

Pour estimer notre modèle VECM, nous allons utiliser les données concernant la RD Congo pour la période qui va de 1976 jusqu'en 2011. Sur la base des critères d'information d'Akaike (AIC) et Schwarz (SC), nous avons retenu le modèle VECM avec p = 1. Le tableau ci-dessous présente les valeurs estimées de l'équation (4) :

*Tableau n° 2. Les résultats de l'estimation du VECM*

| | Variables | ΔINFL | |
|---|---|---|---|
| | | coefficients | t-de student |
| | force de rappel | -0,509 | (-5,2093) |
| | DF | 0,8936 | (2,4928) |
| Long | M2 | 2,6791 | (6,2508) |
| terme | TCE | -0,2637 | (-4,6730) |
| | GDP | -0,1177 | (-1,9752) |
| | ΔDF | 0,4593 | (0,4285) |
| Court | ΔM2 | -0,0542 | (-0,3335) |
| terme | ΔTCE | -1,8357 | (-6,5496) |
| | ΔGDP | 0,0518 | (1,1697) |

*Source: estimation effectuée à partir du logiciel eviews 7.*

$R^2$=0, 78; $R^2$ adj. =0, 73; F-stat. = 16, 56; LM stat = 26, 93 (Prob: 0, 3594); White test (Chi-sq) = 413, 61 (Prob: 0,373); JB= 9, 71 (Prob: 0, 4663); (.) = t- de student.

Globalement, le modèle à correction d'erreur estimé est significatif au regard de la valeur de la statistique F de Fisher qui est de 16,56. La valeur de $R^2$ adj indique que 73% des fluctuations de l'inflation au cours de la période sous étude sont expliquées par les variables du modèle. La valeur du test de Breush-Godfrey indique que les erreurs sont indépendantes au seuil de 5%. L'hypothèse d'absence d'héthéroscédasticité ainsi que celle de normalité des résidus sont acceptées au seuil de 5%, à en croire les probabilités associées aux valeurs statistiques du test white et celui de Jarque-Bera.

Le coefficient de la force de rappel est négatif et statistiquement significatif. Il est compris entre -1 et 0. Ce qui indique que l'inflation s'ajuste à une vitesse de 51% par rapport à son niveau d'équilibre suite à tout choc provenant des variables exogènes. On s'aperçoit donc que le choc se résorbe entièrement au bout d'environ (1/0,51 soit 1,96) deux ans. Les propriétés statistiques de la force de rappel nous permettent ainsi de valider la spécification du modèle à correction d'erreur.

L'identification des indicateurs pertinents d'inflation suppose aussi que soit déterminé en quoi une variable supposée informationnelle explique ou non l'erreur de prévision sur l'inflation. En effet, l'identification des chocs a été reprise en adoptant le schéma d'identification de Cholesky. Le résultat de cette analyse est présenté dans le tableau ci-dessous :

*Tableau n° 3. Analyse de la décomposition de la variance*

| Variance Decomposition of INFL: | | | | | | |
|---|---|---|---|---|---|---|
| Period | S.E. | INFL | DF | M2 | TCE | GDP |
| 1 | 0.680994 | 100.0000 | 0.000000 | 0.000000 | 0.000000 | 0.000000 |
| 2 | 1.394464 | 29.40309 | 1.693995 | 30.34838 | 38.55089 | 0.003646 |
| 3 | 1.766729 | 24.67560 | 2.314871 | 32.22569 | 40.78155 | 0.002285 |
| 4 | 2.037400 | 21.08098 | 2.088448 | 31.46183 | 45.33481 | 0.033936 |
| 5 | 2.189492 | 20.53944 | 1.910258 | **30.21457** | **47.30317** | 0.032564 |
| Cholesky Ordering: INFL DF M2 TCE GDP | | | | | | |

*Source: calculs effectués à l'aide du logiciel E-views 7.*

Les résultats de la décomposition de la variance indiquent que la variance de l'erreur de prévision de l'inflation est due pour 20,5% à ses propres innovations et pour 47,3%, 30,2% et 1,91% à celles respectivement du taux de change, de la masse monétaire et du déficit budgétaire. Nous pouvons dès lors passer à l'interprétation de nos résultats.

### 4. Interpretation des resultats

Deux catégories d'indicateurs d'inflation ont été retenues dans nos analyses, notamment les indicateurs monétaire et financier, d'une part et les indicateurs d'origine réelle.

#### 4.1. Les indicateurs monétaire et financier

Il ressort de nos analyses qu'à court et long terme le coefficient du taux de change présente un signe négatif et est statistiquement significatif au seuil de 1%. Cela traduit une dépréciation persistante de la monnaie nationale face au dollar. On constate alors que toute dépréciation de 1% de la monnaie nationale entraîne une hausse de prix de 1,8% à court terme et 0,26% dans le long terme. Ce résultat s'accorde avec celui obtenu en chine où le taux de change joue aussi un rôle dans la formation des anticipations d'inflation et possède un impact très significatif sur l'inflation (S. GUERINEAU et S. G. JEANNENEY, 2003).

Ce résultat peut s'expliquer par le fait que le pays étant caractérisé par un niveau élevé de dollarisation, la dépréciation du franc congolais face au dollar amène les agents à se dessaisir de la monnaie nationale au profit de la devise, plus précisément le dollar américain, pour préserver un minimum de pouvoir d'achat surtout que le panier de calcul de l'indice des prix à la consommation est essentiellement constitué par les biens et services importés. Il s'en suit que les anticipations à la hausse de la demande des devises accélèrent la dépréciation de la monnaie nationale. Comme les prix des biens et services sont indexés sur l'évolution du taux de change, ils sont à leur tour revus à la hausse ; ce qui finalement entraîne l'inflation.

Le taux de croissance de la masse monétaire présente un coefficient positif et statistiquement significatif seulement à long terme au seuil de 1%. Ce résultat révèle qu'un accroissement de 1% de la masse monétaire se traduit par une augmentation 2,7% du niveau de prix. Il s'ensuit alors qu'à long terme le processus inflationniste en R.D.Congo est expliqué par l'expansion monétaire conformément à la prédiction théorique. Ce résultat corrobore celui obtenu par S.BRANA (1999), J. BAUMGARTNER et al (1996), M. MUHLEISEN (1995) et F. BARARUZUNZA (2009) respectivement, l'Allemagne, la Suède, la Finlande et le Burundi.

#### 4.2. Les indicateurs d'origine réelle

Il ressort de nos résultats que si la relation entre l'inflation et le déficit budgétaire est très faible à court terme, à long terme, la variation du niveau général des prix est fortement expliquée par le déficit budgétaire. Son influence positive est significative au seuil de 5%. Il apparaît donc qu'un accroissement de 1% du déficit budgétaire entraîne une augmentation de 0,89% de l'inflation. Ce résultat trouve sa justification dans le recours accru à la monétisation du déficit budgétaire qui demeure un des facteurs très aggravant de l'inflation en R.D.Congo. Cette conclusion corrobore les résultats obtenus par F. SYLLA et SALL (2007) pour le cas de la Guinée.

Le coefficient du taux de croissance économique est affecté d'un signe négatif à long terme et est statistiquement significatif au seuil de 5%. On s'aperçoit alors qu'en R.D.Congo, l'inflation affecte négativement la croissance économique. Ce résultat traduit l'existence d'un phénomène caractérisé par une stagflation, c'est-à-dire la coexistence d'une inflation persistante et de la baisse du niveau de production.

Ainsi, le coût de l'inflation, c'est-à-dire la perte de croissance induite par un taux d'inflation élevé est extrêmement fort. On s'aperçoit alors que les fluctuations récurrentes de l'inflation ont généré un taux d'inflation moyen très élevé jusqu'à nuire l'économie en ne favorisant pas l'investissement et l'épargne conformément à ce que pensent C. T. NDIAYE et M. A. KONTE, (2012).

Il ressort donc nettement de nos analyses que les indicateurs les plus pertinents d'inflation sont tant d'origine monétaire que réelle. Cette conclusion est validée par les résultats de l'analyse de la décomposition de variance qui place au premier plan le taux de change et la masse monétaire comme des indicateurs les plus pertinents. Ces résultats dégagent de ce point de vue deux enseignements en termes de la conduite de la politique monétaire à court et à long terme.

## 5. Les implications en termes de politique economique

Les résultats de nos analyses placent au premier plan le taux de change et la masse monétaire comme des indicateurs d'inflation les plus pertinents en R.D.Congo. De ce point de vue, deux enseignements en termes de la conduite de la politique monétaire peuvent être dégagés. Plus concrètement, nous allons donc apprécier la stratégie qui doit être mise en place par les autorités monétaires, au travers le contrôle du taux de change d'une part et de la masse monétaire ou du ciblage d'inflation d'autre part, pour stabiliser les anticipations des agents à court et à long terme dans le but de renforcer la crédibilité de la politique monétaire.

### 5.1. L'ancrage nominal par le taux de change

Les enseignements de nos analyses révèlent que les fluctuations du taux de change ont une incidence considérable sur l'inflation non seulement à court terme, mais aussi à long terme. Pour cette raison, c'est la politique d'ancrage nominal du franc congolais au dollar américain qui apparaît plus pertinente pour orienter et stabiliser les anticipations des agents.

Le mécanisme de transmission de la dépréciation du taux de change sur l'inflation en R.D.Congo est donc le suivant : les ménages étant rémunérés en monnaie nationale, toute augmentation inattendue des dépenses publiques financée par la création monétaire diminue la confiance du public dans la devise nationale en faveur du dollar américain ; ce qui entraîne par ricochet une dépréciation du franc congolais qui, à son tour, alimente l'inflation en raison des répercussions d'une hausse des prix des importations[9]. On s'aperçoit donc qu'à court et long terme l'influence du taux de change dans le processus de stabilisation des anticipations est plus grande en R.D.Congo tout comme dans la plupart des pays en développement, à en croire J. P. POLLIN, (2008) et J. B. DESQUILET, (2011).

Le taux de change apparaît comme un signal pertinent et clair pour les agents économiques. En effet, dans une petite économie ouverte, contrôler l'évolution du taux de change permet d'agir directement sur le pouvoir d'achat, qui devrait, à son tour, assurer la stabilité des prix pendant une période raisonnable de temps. Il est attrayant, parce qu'il est facile de l'appliquer, facile pour le public de le comprendre et dans ce cas crédible. Le taux de change constitue une cible contrôlable et permet de soutenir les autorités publiques dans leur engagement en faveur d'un effort de stabilisation et aider les agents privés à coordonner leurs décisions de formation des prix et les anticipations autour d'une cible d'inflation réduit (J. B. DESQUILET, 2011).

---

[9] Comme la quasi-totalité de biens et services pour sa consommation viennent de l'extérieur et que ses exportations reposent essentiellement sur les matières premières à l'état brut dont le pays ne contrôle même pas les prix, l'économie congolaise ne profite pas de gain de compétitivité conformément à la condition de MARSHALL-LERNER.

L'idée que sous-tend l'ancrage nominal du taux de change vis-à-vis du dollar consiste ici à construire la crédibilité de la politique monétaire en s'appuyant sur la réputation d'une banque centrale étrangère, considérée comme plus vertueuse. Le caractère radical d'une telle règle est de nature à provoquer un choc stabilisant sur les anticipations dans des économies fortement inflationnistes comme c'est le cas de l'économie congolaise. Par sa rigidité, surtout lorsqu'elle s'accompagne de la mise en place d'une caisse d'émission, un « currency board »[10], elle peut être une réponse appropriée dans des situations extrêmes (P. DEMPERE et C. QUENAN, 2000).

En effet, dans les pays en transition frappés par le phénomène d'hyperinflation, le mérite de la stabilisation fondée sur le taux de change par rapport à l'objectif de croissance de la masse monétaire ne fait aucun doute si le pays dispose des réserves de change suffisantes pour soutenir le système de change fixe. Les pays en développement d'Amérique latine ou en transition d'Europe centrale et orientale ont largement favorisé l'ancrage nominal par le taux de change. Cette stratégie fournit une règle automatique pour la conduite de la politique monétaire permettant de résoudre le problème d'incohérence temporelle (F. MISHKIN, 2007).

Même si l'ancrage nominal du taux de change au dollar américain apparaît la meilleure stratégie en raison évidemment du niveau élevé de la dollarisation, le suivi de cet objectif peut être sous optimal pour atteindre l'objectif de stabilité des prix. Comme l'économie congolaise fonctionne déjà sous un régime de change flottant depuis 1983, amarré le franc congolais au dollar, cela suppose l'adoption d'un régime plus ou moins fixe. Encore est-il que même si les autorités nationales décident de revenir sur le change fixe, l'ancrage nominal par le taux de change peut aussi présenter beaucoup de limites.

Le ciblage du taux de change limite considérablement la politique monétaire puisqu'elle n'intéresse que le taux de change et restreint la capacité de la banque centrale à réagir aux chocs spécifiques qui frappent l'économie nationale. Elle est difficilement tenable à terme sauf lorsqu'il existe une véritable coopération entre l'économie d'ancrage et celle qui choisit de fixer sa parité. Lorsque le pays recourt à cette stratégie, il perd ipso facto l'indépendance de sa politique monétaire, puisque toutes les décisions, en ce domaine, doivent être essentiellement orientées en fonction de la stabilisation de change : dans le cas d'une politique de caisse d'émission, la création monétaire est strictement déterminée par les entrées et sorties de réserves (E. CROCE et M. S. KHAN, 2000 ; J. P. POLLIN, 2008 ; J. B. DESQUILET, 2011).

L'abandon probablement définitif de l'indépendance de la politique monétaire, du rôle de prêteur en dernier ressort de la part de la Banque centrale et la perte du seigneuriage constitue les « coûts » le plus souvent mis au premier plan.

Le principal inconvénient de cette stratégie est la perte par les autorités de tout contrôle sur leur politique monétaire, devenue désormais dépendante de la Banque Centrale de la monnaie d'ancrage. En effet, le choix d'une autorité monétaire externe comme point de référence implique la subordination de la politique monétaire. La Banque centrale nationale n'aura plus la possibilité de « battre monnaie ». Or, c'est là l'attribution à partir de laquelle elle peut exercer la fonction de prêteur de dernier ressort. Celle-ci est essentielle pour le système bancaire car elle permet de stopper une crise bancaire en fournissant un supplément de crédit. Plus encore, le fait d'ancrer la parité de sa devise à celle d'un autre pays condamne à subir (ou à importer) les chocs réels, monétaires ou financiers, qui touchent ce pays, et donc son taux de change (J. P. POLLIN, 2008).

---

[10] Le currency board, ou caisse d'émission, est caractérisé par trois éléments : un taux de change fixe avec la monnaie de rattachement, la convertibilité automatique et l'engagement crédible des autorités monétaires. Concrètement, il s'agit de garantir la base monétaire par les réserves de change. Les objectifs en sont la restauration de la crédibilité économique, la lutte contre l'inflation et la réduction du niveau moyen des taux d'intérêt domestiques.

Aussi, le système de « currency board », par exemple, est peu adapté aux économies très vulnérables aux chocs extérieurs et à celles qui ont un secteur bancaire affaibli comme c'est le cas de la R.D.Congo. En effet, la politique monétaire n'existe plus dans le sens où la création monétaire ne dépend que de la capacité de l'économie nationale à accroître ses réserves en devises. Dès lors, il n'est plus possible de soutenir un secteur bancaire défaillant (rôle du prêteur en dernier ressort) et d'adapter la masse monétaire aux données conjoncturelles. La politique budgétaire doit elle-même devenir prudente, puisque le financement monétaire du déficit budgétaire n'est plus envisageable. Le renoncement à la souveraineté monétaire peut aller jusqu'à l'abandon d'une monnaie nationale.

En dépit des limites que présentent le ciblage du taux de change, F. MISHIKIN (2007) pense que la perte d'indépendance monétaire découlant d'une politique d'objectif de change est sans doute moins coûteuse pour les pays en développement. Elle peut même être avantageuse, car ces économies ont souvent tout intérêt à amarrer leur monnaie sur celle d'un autre pays au lieu de poursuivre une politique monétaire autonome. C'est sans doute la raison pour laquelle nombreux pays émergents, adoptent un ciblage du taux de change. Les résultats de nos analyses plaident en faveur de cette stratégie pour le cas de la R.D.Congo.

### 5.2. Non pertinence du ciblage monétaire

La Banque Centrale du Congo se rallie à une politique de ciblage des agrégats monétaire pour assurer la stabilité des prix. Cependant, la pertinence de cet objectif laisse à désirer. La faible influence de la monnaie sur l'inflation à court terme fait perdre à la masse monétaire son statut d'objectif intermédiaire pertinent pour assurer la stabilité des prix en R.D.Congo. Ce constat corrobore la conclusion de J. L. NKOULOU NKOULOU (2010) pour le cas de la zone CEMAC. La masse monétaire est donc devenue un simple indicateur avancé de l'inflation en raison de sa forte influence sur les prix à long terme.

Par ailleurs, la remise en cause de l'hypothèse de contrôlabilité de la masse monétaire est aussi validée par le niveau élevé de la dollarisation qui caractérise l'économie congolaise. Il apparaît difficile de déterminer avec exactitude la masse monétaire en circulation et donc des cibles dans la gestion de la politique monétaire dans une économie dollarisée. Dans pareille situation, le contrôle de la base monétaire dont une partie est composée de billets et de dépôts en devises étrangères est rendu moins efficace. Les autorités monétaires se trouvent ainsi dans l'impossibilité de gérer une monnaie nationale devenue incontrôlable. La cible d'agrégat monétaire ne constitue pas un signal clair permettant de coordonner les anticipations inflationnistes surtout que les agents n'ont pas confiance en la monnaie nationale.

Par ailleurs, comme l'économie de la R.D.Congo fonctionne déjà sous un régime de change flottant et que la cible monétaire apparaît moins pertinent pour assurer la stabilité des prix, l'éventualité de basculer vers un régime de ciblage d'inflation est envisageable pour maintenir la stabilité des prix. Dans ce cas, ce sont les prévisions d'inflation qui pourront jouer le rôle d'objectif intermédiaire de la politique monétaire.

### 5.3. Vers le ciblage d'inflation

Le ciblage d'inflation offre un cadre simple et prévisible à la conduite de la politique monétaire qui permet de canaliser les anticipations d'inflation et de les orienter à la baisse à travers le renforcement de la crédibilité et de la transparence de la banque centrale. Il autorise en plus une certaine flexibilité, notamment en cas de choc exogène ou endogène, que l'ancrage du taux de change ne permet pas et offre par conséquent une alternative satisfaisante à la régulation des agrégats monétaires comme objectif intermédiaire (S. CHAUVIN et O. BASDEVANT, 2006).

En raison des conséquences énormes de l'inflation sur le bien-être social en R.D.Congo pendant la période sous étude notamment le ralentissement de la croissance économique, la réduction de l'attractivité de l'économie et donc de la compétitivité des entreprises nationales, il serait souhaitable que les autorités monétaires congolaises de connivence avec le gouvernement annoncent de manière explicite l'objectif quantitatif d'inflation, sous forme ponctuelle ou à travers une fourchette cible, qui serait comptable avec un niveau de croissance durable[11] (M. NORMANDIN, 2008 ; Z. FTITI et J. F. GOUX, 2011).

Par une simple manipulation statistique, nous avons essayé de déterminer le niveau d'inflation compatible avec la croissance économique en R.D.Congo. Les résultats sont donc présentés dans le tableau ci-dessous :

*Tableau n° 4. 1. Statistique descriptive de l'évolution conjointe du taux d'inflation et du taux de croissance économique par période.*

|         | taux d'inflation | | | taux de croissance | | |
|---------|-----------|-----------|-----------|-----------|-----------|-----------|
|         | 1976-1990 | 1991-2001 | 2002-2011 | 1976-1990 | 1991-2001 | 2002-2011 |
| moyenne | 63,33     | 3133,115  | 20,27     | 0,14      | -5,19     | 5,85      |
| médiane | 68,95     | 513,91    | 17,12     | 0,47      | -4,27     | 6,21      |
| maximum | 104,1     | 23773     | 46,1      | 5,54      | 0,7       | 7,79      |
| minimum | 24        | 29        | 3,99      | -6,6      | -13       | 2,8       |

*Sources : nos calculs à partir des données de la Banque Mondiale*

Sans être trop descriptif, nous allons juste nous intéresser à la période qui va de 2002 à 2011 pour déterminer la cible d'inflation. Le choix de cette période est justifié par les performances économiques que la R.D.Congo a enregistrées depuis l'année 2002 après plus d'une décennie marquée par l'hyperinflation et la récession. Il ressort alors de nos calculs que pour maintenir la croissance économique à un taux médian de 6,21 %, il faudra au moins maintenir l'inflation au taux médian de 17,12%. Ce taux pourra varier dans une fourchette de [3,99/17,17%]. En effet, cette cible n'est pas loin de celle adoptée par le Ghana [14,5+/-1] en 2007[12]. Sa détermination nécessite cependant une concertation entre les autorités monétaires et budgétaires et doit faire l'objet d'une analyse rigoureuse.

Ainsi, faute d'un consensus sur le canal prédominant à travers lequel la politique monétaire opère au sein de l'économie pour atteindre son objectif de stabilité des prix, c'est la prévision de l'inflation qui doit tenir lien d'objectif intermédiaire et non pas la masse monétaire ou le taux de change. La banque centrale doit donc veiller sur les indicateurs informationnels reconnus comme étant aptes à prévenir l'inflation pour parvenir à contrecarrer les tensions inflationnistes avant qu'elles ne se concrétisent. L'intervention de la banque centrale sur le marché monétaire va viser dans ce cas à ramener le taux d'inflation anticipé proche de l'inflation ciblée (P. R. AGENOR, 2008 ; L. GREGORY, 2003).

---

[11] En plus de l'annonce d'une cible d'inflation sous forme d'une fourchette ou d'une valeur ponctuelle aussi bien que l'horizon d'engagement, ce régime suppose aussi : 1) l'engagement institutionnel envers la stabilité des prix comme objectif prioritaire de la politique monétaire ; 2) la stratégie utilisant une information vaste, non limitée aux agrégats monétaires et au taux de change pour décider des modifications des instruments chaque fois que l'inflation anticipée diffère de la cible ; 3) la transparence accrue (communication vers le public et les marchés sur les plans, objectifs, et décisions des autorités monétaires) ; 4) la responsabilité accrue de la banque centrale dans la réalisation des objectifs.

[12]

Ces indicateurs comportent toutes choses restant égales par ailleurs la masse monétaire, le taux de change et la consommation publique. L'adoption de cette stratégie apporte une solution pour mieux éclairer aussi bien la Banque Centrale que le grand public sur l'évolution en temps réel de l'objectif final de la politique monétaire. A la différence de la cible intermédiaire de monnaie ou de taux de change, la poursuite de la prévision d'inflation permet une meilleure flexibilité lorsque la Banque centrale est tenue de veiller sur la stabilité du produit à court terme en plus de la stabilisation des prix à long terme (A. LAYOUNI, 2007).

En dépit de problèmes structurels auxquels font face les pays en développement, en l'occurrence la R.D.Congo, le ciblage de l'inflation semble prometteur. Il offre un certain nombre d'avantages opérationnels et oblige les responsables de la politique économique à approfondir les réformes, à accroître la transparence et à améliorer la politique budgétaire. Il leur ouvre également la perspective d'une convergence vers les niveaux internationaux d'inflation (E. CROCE et M. S. KHAN, 2000).

## 6. Conclusion

En somme, notre travail a porté sur *les indicateurs avancés de l'inflation en RD Congo*. En abordant ce thème, nous avons cherché à apprécier la pertinence de la masse monétaire utilisée par la Banque Centrale du Congo[13] comme cible intermédiaire et d'identifier le cas échéant d'autres indicateurs performants qui peuvent être suivis par les autorités monétaires dans la conduite de leur action. Ce qui nous a conduit à formuler notre question de recherche en ces termes : quels sont les indicateurs pertinents pour la conduite de la politique monétaire en RD Congo ?

Pour répondre à cette question, le modèle développé par J. H. STOCK et M. W. WATSON (1999) tel que synthétisé par S. KOZICKI (2001) et adapté aux spécificités de la R.D.Congo nous a servi de cadre de référence. Nous avons recouru à l'approche de Johansen pour déterminer le nombre des vecteurs de cointégration et justifier l'utilisation du modèle à correction d'erreurs.

Les résultats obtenus révèlent qu'en R.D.Congo les indicateurs d'inflation les plus pertinents sont tant d'origine monétaire que réelle. Ce sont donc les chocs d'origine monétaire qui affectent beaucoup plus l'économie congolaise. Cette conclusion est validée par les résultats de l'analyse de la décomposition de variance qui placent au premier plan le taux de change, la masse monétaire et la consommation publique comme des indicateurs les plus pertinents.

Afin d'atteindre son objectif de stabilité des prix et soutenir une croissance économique durable, l'objectif intermédiaire de la Banque Centrale fondé sur le contrôle de la masse monétaire paraît moins pertinent. Compte tenu du niveau élevé de la dollarisation, la banque centrale pourra soit adopter la stratégie d'ancrage nominal du taux de change vis-à-vis du dollar, ce qui suppose le retour du change fixe ; soit adopter une stratégie de ciblage d'inflation pour restaurer la crédibilité de la politique monétaire.

---

[13] Dès lors que l'objectif intermédiaire, masse monétaire, crédit à l'économie ou taux de change, suivi par la banque centrale contient toute l'information utile à la prévision de l'inflation, utiliser un objectif intermédiaire est équivalent à cibler la prévision de l'inflation (B. S. BERNANKE et F. S. MISHKIN, 1997).

Pour ce faire, les autorités nationales devraient par ailleurs promouvoir un mécanisme de financement des investissements fondé sur l'accroissement d'une épargne stable et volontaire et non sur la création d'actifs monétaires. Ce qui suppose une discipline budgétaire et de surcroît renforcer l'indépendance de la banque centrale.

Nous n'avons pas la prétention de présenter les conclusions de nos analyses comme des certitudes et des vérités implacables. Nous estimons néanmoins qu'il serait souhaitable d'approfondir nos analyses en menant une étude sur *la règle monétaire optimale dans une petite économie ouverte fortement « dollarisée » ; le cas de la R.D.Congo par exemple.*

## BIBLIOGRAPHIE